\documentclass[
    ,final            
    ,final            
  ,sort&compress
  ]
  {aipproc}
\usepackage{amsmath}
\usepackage{units}
\usepackage{graphicx}
\usepackage{braket}
\usepackage[normalem]{ulem}
\usepackage{booktabs}
\usepackage{tabularx}

\newcolumntype{Y}{>{\centering\arraybackslash}X}
 
\layoutstyle{8x11double}
\begin{document}
\setlength{\AIPhlinesep}{0pt}
\setlength{\abovedisplayskip}{6pt}
\setlength{\belowdisplayskip}{6pt}
\setlength{\abovedisplayshortskip}{0pt}
\setlength{\belowdisplayshortskip}{4pt}
 
\addtolength{\textfloatsep}{-18pt}
\addtolength{\dbltextfloatsep}{-12pt}
 
\title{Student Difficulties with Quantum Mechanics Formalism}
 
 
\author{Chandralekha Singh}{
  address={Department of Physics and Astronomy, University of Pittsburgh, Pittsburgh, PA 15260}
}
 
\date{2 February 2016}

\begin{abstract}
We discuss student difficulties in distinguishing between the physical space and Hilbert space and
difficulties related to the Time-independent Schroedinger equation and measurements in quantum mechanics.
These difficulties were identified by administering written surveys and by conducting individual interviews with students.
\end{abstract}

\maketitle

\section{Introduction}
\vspace*{-.08in}

Here, we describe the difficulties with the formalism of quantum mechanics identified by 
administering written surveys to eighty-nine advanced undergraduates and more than two hundred graduate students from 
seven different universities.
In the written surveys, students were asked to explain their reasoning.
We also conducted 
individual interviews using a think aloud protocol
with a subset of students to better understand the rationale for their responses.
During the semi-structured interviews, students were asked to verbalize their thought processes while they answered
qualitative questions posed to them. Students were not interrupted unless they remained quiet for a while. 
In the end, we asked them for clarifications of the issues they had not made clear earlier. 
Below we discuss some of the difficulties with the formalism.

\vspace*{-.15in}
\subsection{Difficulty Distinguishing between the Physical Space and Hilbert Space}
\vspace*{-.05in}

In quantum theory, one must interpret the outcome of real experiments performed in three dimensional (3D) space by
making connection with an abstract Hilbert space (state space) in which the wavefunction lies. The
physical observables that one measures in the laboratory correspond to Hermitian operators in the Hilbert space whose
eigenstates span the space. Knowing the initial wavefunction and the Hamiltonian of the system allows
one to determine the time-evolution of the wavefunction unambiguously and the measurement postulate can be used to
determine the possible outcomes of individual measurements and ensemble averages (expectation values).

It is difficult for many students to distinguish between vectors in the 3D laboratory space and Hilbert space. For example,
$S_x$, $S_y$ and $S_z$ denote the orthogonal components of the spin angular momentum vector of an electron in the 3D space, each
of which is a physical observable that can be measured in the laboratory. However, the Hilbert space corresponding to the
spin degree of freedom for a spin-1/2 particle is two dimensional (2D). In this Hilbert space, $\hat S_x$, $\hat S_y$ and $\hat S_z$
are operators whose eigenstates span the 2D space.
The eigenstates of $\hat S_x$ are vectors which span the 2D space and are orthogonal to each other
(but not orthogonal to the eigenstate of $\hat S_y$ or $\hat S_z$).
Also, $\hat S_x$, $\hat S_y$ and $\hat S_z$ are operators and
{\it not} orthogonal components of a vector in the 2D space. If the electron is in a magnetic field with the gradient
in the $z$ direction in the laboratory (3D space) as in a Stern-Gerlach experiment, the magnetic field is a vector
in 3D space but not in the 2D space. It does not make sense to compare vectors in the 3D space with
vectors in the 2D space as in statements such as ``the magnetic field gradient is perpendicular to the eigenstates of $\hat S_x$".
Unfortunately, these distinctions are difficult for students to make and such difficulties were frequently observed in response
to the survey questions and during individual interviews. This difficulty is discussed below in the context of
a two part question related to the Stern-Gerlach experiment:

Question: Notation: $\vert \uparrow_z \rangle$ and  $\vert \downarrow_z \rangle$ represent the orthonormal eigenstates of
$\hat S_z$ ($z$ component of the spin 
of the electron). SGA is an abbreviation for a
Stern-Gerlach apparatus.  The electron is in the SGA for an infinitesimal time.
Ignore the Lorentz force on the electron.

(a) A beam of electrons propagating along the $y$ direction (into the page) in spin state $(\vert \uparrow_z \rangle+\vert \downarrow_z \rangle)/\sqrt{2}$
is sent through an SGA with a vertical magnetic field gradient in the $-z$ direction. Sketch the
electron cloud pattern that you expect to see on a distant phosphor screen in the x-z plane. Explain your reasoning.

(b) A beam of electrons propagating along the $y$ direction (into the page) in spin state $\vert \uparrow_z \rangle$
is sent through an SGA with a horizontal magnetic field gradient in the $-x$ direction.
Sketch the electron cloud pattern that you expect to see on a distant phosphor screen in the x-z plane. Explain your reasoning.

In question (a), students have to realize that the magnetic field gradient in the -z direction would exert a force
on the electron due to its spin angular momentum and one should observe two spots on the phosphor screen due to the splitting
of the beam along the z direction corresponding to electron spin component in
$\vert \uparrow_z \rangle$ and $\vert \downarrow_z \rangle$ states.
All responses in which students noted that there will be a splitting along the z direction
were considered correct even if they did not explain their reasoning.
Only $41\%$ of the students provided the correct response.
Many students thought that there will only be a single spot on the phosphor screen as in these typical survey responses:
{\it 
\begin{itemize}
\item SGA will pick up the electrons with spin down since the gradient is in -z direction.
The screen will show electron cloud only in $-z$ part.
\item All of the electrons that come out of the SGA will be spin down with expectation value
$-\hbar/2$ because the field gradient is in $-z$ direction.
\item  Magnetic field is going to align the spin in that direction so most of the electrons will align along -z direction. We may
still have a few in the +z direction but the probability will be very small.
\end{itemize}
}
In the interviews, students were often confused about the origin of the force on the particles and
whether there should be a force on the particles at all as they pass through the SGA.

Question (b) is more challenging than (a) because students have to realize that the
eigenstate of $\hat S_z$, $\vert \uparrow_z \rangle$ can be
written as a linear superposition of the eigenstates of $\hat S_x$, i.e.,
$\vert \uparrow_z \rangle=(\vert \uparrow_x \rangle + \vert \downarrow_x \rangle) /\sqrt{2}$.
Therefore, the magnetic field gradient in the $-x$ direction will
split the beam along the x direction corresponding to the electron spin components
in $\vert \uparrow_x \rangle$ and $\vert \downarrow_x \rangle$ states and cause two spots on the phosphor screen.
Only $23\%$ of the students provided the correct response.
The most common difficulty was assuming that since the spin state is $\vert \uparrow_z \rangle$,
there should not be any splitting as in the examples below:

{\it 
\begin{itemize}
\item Magnetic field gradient cannot affect the electron because it is perpendicular to the wavefunction.
\item Electrons are undeflected or rather the beam is not split since $\vec B$ is perpendicular to spin state.
\item The direction of the spin state of the beam of electrons is $y$, and the magnetic field gradient is in the $-x$ direction. The two directions have an angle $90^0$, so the magnetic field gradient gives no force to electrons.
\item  With the electrons in only one measurable state, they will experience a force only in one direction upon interaction with $\vec B$.
\end{itemize}
}
Thus, many students explained their reasoning by claiming
that since the magnetic field gradient is in the $-x$ direction but
the spin state is along the $z$ direction, they are orthogonal to each other,
and therefore, there cannot be any splitting of the beam. Student responses suggest that
they were incorrectly connecting the direction of magnetic field in the 3D space
with the ``direction" of state vectors in the Hilbert space.
Several students in (b) drew a monotonically increasing function.
One interviewed student drew a diagram of a molecular orbital with four lobes
and said ``this question asks about the electron cloud pattern due to spin...I am wondering what the spin part of the wavefunction
looks like." Then he added,
``I am totally blanking on what the plot of $\vert \uparrow_z \rangle$ looks like otherwise
I would have done better on this question". From responses such as these it appears that the
abstract nature of spin poses special problems in teaching quantum physics.

Compared to question (a), many more students in (b) thought that there will be only one spot on the screen,
but there was no consensus on the direction of deflection despite the fact that students were asked to ignore the Lorentz force. Some students drew
the spot at the origin, some showed deflections along the positive or negative x direction, some along the positive or negative z direction.
They often provided interesting reasons for their choices.
Some students who drew two spots were confused about the direction in which the magnetic field gradient will cause the splitting of the beam.
Thirteen percent of the students (including questions (a) and (b)) drew the splitting of the beam in the wrong direction (along the x axis
in (a) and along the z axis in (b)).
One interviewed student who drew it in the wrong direction said, ``I remember doing this recently and I know there is some splitting
but I don't remember in which direction it will be."

Students were posed another question involving $\vec S \cdot \hat n$ where 
$\hat n$ is a general unit vector pointing in an arbitrary direction in the physical three dimensional space. This dot product is
a scalar product between two vectors in the physical space. Students were given that for spin 1/2, a state $\chi$ goes to $\chi^\prime$
via $\chi^\prime=e^{i(\vec S \cdot \hat n) \phi/\hbar} \chi$ where $e^{i(\vec S \cdot \hat n) \phi/\hbar}$ effects a rotation through
angle $\phi$ about the axis $\hat n$. They were 
asked to construct a $2 \times 2$ matrix representing rotation by
$\phi=\pi$ about the $x$ axis and show that it converts $\chi_+$ to $\chi_-$ ($\chi_\pm$ are the eigenstates of $\hat s_x$ with eigenvalues $\pm \hbar/2$).
Written responses and interviews suggest that one major difficulty was that many students were confused
about whether $\vec S \cdot \hat n$ (which can be written as $s_x n_x+s_y n_y+s_z n_z$) is a dot product in the physical space or Hilbert space. 
Students were not clear about the fact that in a
Hilbert space, the possible states of the system are the vectors and the inner products of these states are the scalar products.
Similar confusion between physical space and Hilbert space
were observed in the context of questions posed in surveys and interviews about a one-dimensional infinite square well.
The physical space for this problem is one-dimensional (e.g., in a quantum wire) but the Hilbert 
space is infinite dimensional.

\vspace*{-.15in}
\subsection{Difficulties Related to Time-independent Schroedinger Equation and Measurement}
\vspace*{-.05in}

One of the questions on the survey asks students to consider the following statement: ``By definition, the Hamiltonian
acting on any allowed state of the system $\psi$ will give the same state back, i.e., $H\psi=E \psi$, where $E$ is the energy
of the system." Students were asked to explain why they agree or disagree with this statement.
We wanted students to disagree with the statement and note that it is only true if $\psi$ is a stationary state.
In general, $\psi=\sum_{n=1}^{\infty} C_n \phi_n$, where $\phi_n$ are the stationary
states and $C_n=\langle \phi_n \vert \psi \rangle$. Then, $\hat H \psi=\sum_{n=1}^{\infty} C_n E_n \phi_n \ne E \psi$. 
For this question, just writing down ``disagree" was not enough for the response to be counted correct. 
Students had to provide the correct reasoning.
Only $29\%$ of the students provided the correct response with correct reasoning.
Thirty-nine percent of students wrote (incorrectly) that the statement is unconditionally correct. 
Typically, these students were reasonably confident of their responses as can be seen from these examples:
{\it
\begin{itemize}
\item Agree, this is a fundamental postulate of quantum mechanics which is proved to be highly exact until present.
\item Agree. Hamiltonian does not alter the state of the system.
\item Agree. Hamiltonians give back physical observables energy. It is an observable and real.
\end{itemize}
}

In response to this question, $10\%$ of students agreed with the statement as long as the Hamiltonian is not time-dependent.
They often claimed (incorrectly) that if $\hat H$ is not time-dependent,
the energy for the system is conserved so $\hat H \psi=E \psi$ must be true.
The following are typical examples:
{\it
\begin{itemize}
\item Agree, if the potential energy does not depend on time.
\item Agree but only if the energy is conserved for this system.
\item Agree because energy is a constant of motion.
\item Agree if it is a closed system since $H$ is a linear operator and gives the same state back multiplied by the energy.
\end{itemize}
}
While the energy is conserved if the Hamiltonian is time-independent, $\hat H \psi=E \psi$ need not be true. For example,
if the system is in a linear superposition of stationary states, $\hat H \psi\ne E \psi$ although the energy is conserved.

Eleven percent ($11\%$) of the students answering this question believed (incorrectly) that any statement involving a
Hamiltonian operator acting on a state is a statement about the measurement of energy.
Some of these students who (incorrectly) claimed that $\hat H \psi=E \psi$ is a statement about energy measurement agreed
with the statement while others disagreed.
Those who disagreed often claimed that $\hat H \psi=E_n \phi_n$
because as soon as $\hat H$ acts on $\psi$, the wavefunction will
collapse into one of the stationary states $\phi_n$ and the corresponding energy $E_n$ will be obtained.
The examples below are typical of students with this misconception:
{\it 
\begin{itemize}
\item Agree. If you make a measurement of energy by applying $H$ to a state of an electron in hydrogen atom you will get the energy.
\item Disagree. Hamiltonian acting on a state (measurement of energy) will return an energy eigenstate.
\item Disagree. Quantum measurements will perturb the system so that it jumps into an eigenstate after measurement.
\item Disagree. If it is a mixed state, the measurement of energy will force it to end up with some base state.
\item Disagree. When $\Psi$ is a superposition state and $\hat H$ acts on $\Psi$, then $\Psi$ evolutes to one of the 
$\Psi_n$ so we have $\hat H \Psi=E_n \Psi_n$.
\end{itemize}
}
Interviews and written reasonings suggest that these students believed that
the measurement of a physical observable in a particular state
is achieved by acting with the corresponding operator on the state.
The incorrect notions expressed above are over-generalizations of the fact that
{\it after} the measurement of energy, the system is in a stationary state so $\hat H \phi_n=E_n \phi_n$.

Individual interviews related to this question suggest that some students believed that
whenever an operator $\hat Q$ corresponding to a physical observable $Q$ acts on {\it any} state $ \psi$, it will
yield the corresponding eigenvalue $\lambda$ and the same state back, i.e., $\hat Q \psi=\lambda \psi$.
Some of these students were over-generalizing their ``$\hat H \psi=E \psi$" reasoning and attributing
$\hat Q \psi=\lambda \psi$ to the measurement of an observable $Q$.
Before over-generalizing to any physical observables, these students often agreed with the $\hat H \psi=E \psi$
statement with arguments such as ``the Hamiltonian is the quantum mechanical operator which corresponds to the
physical observable energy" or ``if $H$ did not give back the same state it would not be a hermitian operator
and therefore would not correspond to an observable". Of course, $\hat Q \psi \ne \lambda \psi$ unless $\psi$
is an eigenstate of $\hat Q$ and in general
$\psi=\sum_{n=1}^{\infty} D_n \psi_n$, where $\psi_n$ are the eigenstates of $\hat Q$
and $D_n=\langle \psi_n \vert \psi \rangle$. Then, $\hat Q \psi=\sum_{n=1}^{\infty} D_n \lambda_n \psi_n$ (for
observable with a discrete eigenvalue spectrum).

Many students believed that 
even when answering questions related to the probability of different possible outcomes for the
measurement of an observable other than energy, the wavefunction should be expanded in terms of the energy eigenfunctions
and the absolute square of the expansion coefficients will give the probability of
measuring different values of that observable. In contrast, the wave function should be expanded
in terms of the eigenfunctions of the operator corresponding to the physical observable to be measured and
the absolute square of the expansion coefficients then will give the probability of measuring different possible values
of that observable. Student's belief that the energy eigenfunctions are always the ``preferred" basis vectors is not
surprising because quantum mechanics courses often exclusively focus on solving time-independent Schroedinger equation
to find the energy eigenfunctions and eigenvalues. Also, for 
questions related to the time-development of the wavefunction one must expand the wavefunction in terms of the energy eigenfunctions.

Moreover, students often believed that successive measurements of continuous variables, {\it e.g.}, position, produce
``somewhat" deterministic outcomes whereas successive  measurements of
 discrete variables, {\it e.g.}, spin, can produce very different outcomes.
In an interview, one student began with a correct statement: ``{\em if you measure (an observable) Q, the system will collapse into an
eigenstate of that operator. Then, if you wait for a while the measurement will be different}". But then he added incorrectly:
``{\em if Q has a continuous spectrum then the system would gently evolve and the next measurement won't be very different
from the first one. But if the spectrum of eigenvalues is discrete then you will get very different answers even if you did
the next measurement after a very short time}". When the student was asked to elaborate, he said:
``{\em For example, imagine measuring the position of an electron. It is a continuous function so the time dependence is gentle and
after a few seconds you can only go from A to its neighboring point. [Pointing to an x vs. t graph that he sketches on the
paper]...you cannot go from this place to this without going through this intermediate space}".
When asked to elaborate on the discrete spectrum case, he said:
``{\em...think of discrete variables like spin...they can give you very different values in a short time because the system must
flip from up to down. I find it a little strange that such [large] changes can happen almost instantaneously. But that's what
quantum mechanics predicts...}"
This type of response may also be due to the difficulty in reconciling classical and quantum mechanical ideas; in
classical mechanics the position of a particle is deterministic and can be unambiguously predicted for all times
from the knowledge of the initial conditions and potential energy.

\vspace*{-.16in}
\subsection{Confusion between the Probability of Measuring Position and $\langle x \rangle$}
\vspace*{-.06in}

Born's probabilistic interpretation of the wavefunction can also be confusing for students.
In one question, students were told that immediately after the measurement of energy which yields the first excited state, 
a measurement of the electron position is performed.
They were asked to describe qualitatively the possible values of position they could measure and the probability of measuring them. 
We hoped that students would note that one can measure position values between $x=0$ and $x=a$ (except at $x=0, a/2,a$ where the 
wavefunction is zero), and according to Born's interpretation, $\vert \phi_2(x) \vert^2 dx$ gives the probability of finding the 
particle in a narrow range between $x$ and $x+dx$. Only $38\%$ of the students
provided the correct response. Partial responses were considered correct for tallying purposes
if students wrote anything that was correct related to the above wavefunction, {\it e.g.},
``The probability of finding the electron is highest at $a/4$ and $3a/4$.", ``The probability of finding the electron is
non-zero only in the well", etc.

Eleven percent of the students tried to find the expectation value of position $\langle x \rangle$ instead of the probability of finding the electron
at a given position. They wrote the expectation value of position in terms of an integral involving the wavefunction. Many of them
explicitly wrote that $Probability=(2/a)\int_0^a x$ $sin^2(2\pi x/a) dx$ and
believed that instead of $\langle x \rangle$ they were calculating the probability of measuring the position of electron.
During the interview, one student said (and wrote on paper) that the probability is
$\int x$ $\vert \Psi \vert^2 dx$. When the interviewer asked why $\vert \Psi \vert^2 $ should be multiplied with
$x$ and if there is any significance of $\vert \Psi \vert^2 dx $ alone without multiplying it by $x$, the student said,
``$\vert \Psi \vert^2 $
gives the probability of the wavefunction being at a given position and if you multiply it by $x$ you get the probability of
{\it measuring} (student's emphasis) the position $x$". When the student was asked questions about the meaning of the
``wavefunction being at a given position", and the purpose of the integral and its
limits, the student was unsure. He said that the reason he wrote the integral is because
$x$ $\vert \Psi \vert^2 dx$ without an integral looked strange to him. Similar confusion about probability in classical physics situations have
been found~\cite{whitmann}.

\vspace*{-.22in}
\section{Conclusion}
\vspace*{-.07in}

Instructional strategies that focus on improving student understanding of these concepts should take into account these 
difficulties~\cite{zollman,redish,my}.
We are currently developing and assessing Quantum Interactive Learning Tutorials (QuILT)
suitable for use in advanced undergraduate quantum mechanics courses~\cite{singh}. 

\vspace*{-.23in}
\begin{theacknowledgments}
We are grateful to the NSF for award PHY-0244708.
\end{theacknowledgments}
\vspace*{-.21in}

\vspace*{-.065in}
\bibliographystyle{aipproc}

\end{document}